\documentclass[12pt]{amsart}
\usepackage{amsmath}

\newtheorem{theorem}{Theorem}
\newtheorem{remark}{Remark}

\newcommand{\N}{\mathbb N}
\newcommand{\C}{\mathbb C}
\newcommand{\R}{\mathbb R}

\newcommand{\fii}{\varphi}
\newcommand{\hi}{\mathcal H}
\def\<{\langle}
\def\>{\rangle}
\def\d{{\mathrm d}}
\newcommand{\cal}{\mathcal}

\newcommand{\ket}[1]{\left|#1\right\rangle}  
  
\newcommand{\kb}[2]{\left|#1\right\rangle\left\langle#2\right|}

\begin{document}

\title{The phase representation of covariant phase observables}
\author{Juha-Pekka Pellonp\"a\"a}
\address{Department of Physics, University of Turku, 
20014 Turku, Finland}
\email{juhpello@utu.fi}
\maketitle

\sloppy


\begin{abstract}

Covariant phase observables are obtained by defining simple conditions for
mappings from the set of phase wave functions
(unit vectors of the Hardy space) to the set of phase probability densities.
The existence of phase probability density for any phase wave function, 
the existence of interference effects, and the natural phase shift covariance are 
those simple conditions.
The nonlocalizability of covariant phase observables is proved.

\end{abstract}

\vspace{1cm}


\vspace{1cm}


\vspace{1cm}


Let us start with some
necessary conditions of phase measurements.
From a phase measurement one gets phase statistics, that is,
a probability density $f$ defined on the phase interval, say, $[0,2\pi)$.
Thus, $(2\pi)^{-1}\int_0^{2\pi}f(\theta)\d\theta=1$, $f(\theta)\ge0$, $\theta\in[0,2\pi)$, 
and $f$ can be expanded periodically to the whole real line.
The probability for phase to be in the phase window $X\subseteq[0,2\pi)$ is
$(2\pi)^{-1}\int_X f(\theta)\d\theta$.

In quantum theory, one associates a Hilbert space to the physical system 
under consideration. In the case of a single mode optical field, it is convenient to
choose the number space $\hi_N$ spanned by number states $(|n\rangle)_{n\in\N}$
to describe the system. The number space is isomorphic to the Hardy class $H^2$ on 
the unit disc via the mapping $|n\>\mapsto e^{-in\theta}$. 
Hence, an element $\psi$ of $H^2$ can be represented as a Fourier series
$\psi(\theta)=\sum_{n=0}^\infty d_ne^{-in\theta}$ where 
$\sum_{n=0}^\infty|d_n|^2<\infty$. The Hardy class $H^2$ is a subspace of 
$L^2([0,2\pi))$, the $L^2([0,2\pi))$-space
consisting of all square integrable functions $[0,2\pi)\to\C$ with Fourier-series
$\sum_{n=-\infty}^\infty d_ne^{-in\theta}$ where $\sum_{n=-\infty}^\infty|d_n|^2<\infty$.

The number operator $N=\sum_{n=0}^\infty n\kb n n$ is isomorphic to the
derivative operator $i\partial_\theta$ in $H^2$, and the phase shifter $e^{i\alpha N}$
operates as $\psi(\theta)\mapsto\psi(\theta-\alpha)$ in $H^2$.
It is natural to say that $H^2$ is a {\it phase representation space} for the 
single mode system, and we call (normalized) elements of $H^2$ as 
{\it phase wave functions}.

Suppose that the single mode system is prepared in a (vector) state 
$\psi\in H^2$, $\|\psi\|=1$, and we make 
phase measurements and get a phase probability density $f$.
The canonical distribution is $f(\theta)=|\psi(\theta)|^2$ and 
the trivial (random) distribution is $f(\theta)=1$. 
Next we state general conditions for phase measurements.
\begin{enumerate}
\item {\bf Existence of probability densities.} For any phase wave function
$\psi\in H^2$ there exists a phase probability density $f_{\psi,\psi}:\,[0,2\pi)\to[0,\infty]$.
\item {\bf Existence of interference effects.} For any two phase wave functions
$\psi,\;\fii\in H^2$ there exist complex valued
functions $f_{\psi,\fii}$ and $f_{\fii,\psi}$
defined on $[0,2\pi)$ such that
$$
f_{c_1\psi+c_2\varphi,c_1\psi+c_2\varphi}(\theta)=|c_1|^2 f_{\psi,\psi}(\theta)+
|c_2|^2 f_{\varphi,\varphi}(\theta)+\overline{c_1}c_2f_{\psi,\varphi}(\theta)+
c_1\overline{c_2}f_{\varphi,\psi}(\theta)
$$
for all $c_1,\,c_2\in\C$ such that $\|c_1\psi+c_2\varphi\|=1$.
\item {\bf Phase shift covariance.} If one shifts the phase of a phase wave function $\psi$, 
that is, $\psi(\theta)\mapsto\psi^\alpha(\theta):=\psi(\theta+\alpha)$, 
then the probability density $f_{\psi,\psi}$ 
should only move without deforming its shape. Hence,
$$
f_{\psi^\alpha,\psi^\alpha}(\theta)=f_{\psi,\psi}(\theta+\alpha)
$$
for all $\theta,\,\alpha\in\R$.
\end{enumerate}
Anticipating the below results, we say that the mapping
$(\psi,\fii)\mapsto f_{\psi,\fii}$, which
satisfies Conditions 1, 2, and 3, is a {\it phase observable}.
It follows from Condition 2 that any phase observable has a unique sesquilinear 
extension to $H^2\times H^2$.
From Condition 1 one sees that this mapping
is bounded (in the $L^1$-norm). Thus, from Proposition 2 of \cite{Pe2}
one gets
$$
f_{\psi,\fii}=\sum_{n,m=0}^\infty f_{\eta_n,\eta_m}\overline{a_n}b_m
$$
where $\eta_n(\theta):=e^{-in\theta}$, 
$\psi(\theta)=\sum_{n=0}^\infty a_ne^{-in\theta}$, and
$\fii(\theta)=\sum_{n=0}^\infty b_ne^{-in\theta}$.
Since
$$
\eta^\alpha_n(\theta):=\eta_n(\theta+\alpha)=e^{-in\alpha}\eta_n(\theta)
$$ 
and due to the sesquilinearity
$$
f_{\eta^\alpha_n,\eta^\alpha_m}=e^{i(n-m)\alpha}f_{\eta_n,\eta_m}
$$
it follows from Condition 3 that
$$
f_{\eta_n,\eta_m}(\alpha)=f_{\eta^\alpha_n,\eta^\alpha_m}(0)=e^{i(n-m)\alpha}c_{n,m}
$$
where $c_{n,m}:=f_{\eta_n,\eta_m}(0)$. 
In order that $\theta\mapsto f_{\fii,\fii}(\theta)$ is positive and normalized, 
it is necessary that $(c_{n,m})$ is
positive semidefinite matrix with diagonals equal one, that is,
$(c_{n,m})$ is a {\it phase matrix} \cite{LaPe2}.
Hence, we can write (see Section 8 of \cite{Pe2}) that
$$
f_{\psi,\fii}(\theta)=\sum_{n,m=0}^\infty c_{n,m}e^{i(n-m)\theta}\overline{a_n}b_m
$$
and
$$
\frac{1}{2\pi}\int_X f_{\psi,\fii}(\theta)\d\theta=
\sum_{n,m=0}^\infty c_{n,m}\,\frac{1}{2\pi}\int_Xe^{i(n-m)\theta}\d\theta\,\overline{a_n}b_m
$$
for all (Borel) sets $X\subseteq[0,2\pi)$.

For the canonical distribution, $f_{\psi,\fii}(\theta)=\overline{\psi(\theta)}\fii(\theta)$,
$c_{n,m}=1$ for all $n,\,m$, and for the trivial distribution 
$f_{\psi,\fii}(\theta)=(2\pi)^{-1}\int_0^{2\pi}\overline{\psi(\theta)}\fii(\theta)\d\theta$,
$c_{n,m}=\delta_{n,m}$ (Kronecker's delta).
Note that
the mapping $(\psi,\fii)\mapsto f_{\psi,\fii}$ can be extended to the linear mapping from
the set of trace-class operators on $H^2$ to the set of integrable functions
$L^1([0,2\pi))$ (see Proposition 2 of \cite{Pe2}).
Especially, for any state (i.e., a positive trace-one operator) one
gets a probability density function and, thus, a probability measure.
A mapping from the set of states to the set of probability measures 
(on the outcome space of a measurement) defines a unique normalized positive
operator measure (POM) \cite{OQP} and, thus, it is reasonable to call the mapping 
$(\psi,\fii)\mapsto f_{\psi,\fii}$ an observable.

A phase matrix $(c_{n,m})$ defines a unique POM, 
a {\it (covariant) phase observable} \cite{Ho,LaPe}
$$
E(X)=\sum_{n,m=0}^\infty c_{n,m}\frac{1}{2\pi}\int_Xe^{i(n-m)\theta}\d\theta
\kb n m
$$
for all (Borel) sets $X\subseteq[0,2\pi)$. Thus, using a unitary mapping
$U:\,\cal H_N\to H^2$, $\ket n\mapsto U(\ket n):=\eta_n$, 
$$
\frac{1}{2\pi}\int_ Xf_{\psi,\fii}(\theta)\d\theta=\<\psi|UE(X)U^*\fii\>,\;\;\;\;\psi,\,\fii\in H^2,
$$
which shows that we can understand a phase observable as a phase shift covariant 
POM $E$, or a mapping $(\psi,\fii)\mapsto f_{\psi,\fii}$ satisfying conditions 1, 2, and 3.

Theorem 2 of \cite{Pe2} states that
$$
E(X)=\sum_{n=0}^\infty V_n\,\frac{1}{2\pi}\int_X|\theta)(\theta|\d\theta\,V_n^*
$$
where $|\theta):=\sum_{n=0}^\infty e^{in\theta}\ket n$ is 
the London phase state, and $V_n=\sum_{k=0}^\infty z_{n,k}\kb k k$ are 
bounded operators for which $\sum_{n=0}^\infty|z_{n,k}|^2=1$ for all $k\in\N$.
The mapping $X\mapsto(2\pi)^{-1}\int_X|\theta)(\theta|\d\theta$ is 
the {\it canonical phase observable} \cite{Ho, LaPe2}.
Note that $\|V_n\|\le1$ and thus $\|V_n\psi\|\le\|\psi\|$.
In the case of the canonical phase, $V_0=I$ and $V_n=O$, $n\ge 1$, whereas 
in the case of the trivial phase, $V_n=\kb n n$ for all $n$.
The canonical phase is (up to unitary equivalence) the only phase observable which
is detemined by only one phase state $V_n|\theta)$.
Defining contractions $v_n:=U V_n^*U^*$ we may write 
$$
\frac{1}{2\pi}\int_  Xf_{\psi,\fii}(\theta)\d\theta=\sum_{n=0}^\infty\frac{1}{2\pi}\int_X
\overline{v_n\psi(\theta)}v_n\fii(\theta)\d\theta,\;\;\;\;\psi,\,\fii\in H^2.
$$

For any $s\in\N$ we may define a continuous function 
$(x,y)\mapsto C_s(x,y):=\sum_{n,m=0}^s e^{-inx}c_{n,m}e^{imy}$.
Thus, one gets
$$
\frac{1}{2\pi}\int_Xf_{\fii,\psi}(\theta)\d\theta=\lim_{s\to\infty}\frac{1}{2\pi}\int_X\left[
\frac{1}{2\pi}\int_0^{2\pi}\frac{1}{2\pi}\int_0^{2\pi}\overline{\fii(x)}
C_s(x-\theta,y-\theta)\psi(y)\d x\d y\right]\d\theta.
$$
In the case of the canonical phase we may formally write that
$$
\lim_{s\to\infty}C_s(x,y)\to2\pi\delta_{2\pi}(x)\,2\pi\delta_{2\pi}(y)
$$
where $\delta_{2\pi}$ is the $2\pi$-periodic Dirac delta function\footnote{
Note that formally $\delta_{2\pi}(x)=(2\pi)^{-1}\sum_{n=-\infty}^\infty e^{inx}$ operating
in a suitable subset of $L^2([0,2\pi))$ as an integral ($f\mapsto
\lim_{s\to\infty}\int_0^{2\pi}f(x)(2\pi)^{-1}\sum_{n=-s}^s e^{inx}\d x=f(0)$).
If we restrict the mapping $f\mapsto f(0)$ to the subset of $H^2$ then formally writing
$\delta_{2\pi}(x)=(2\pi)^{-1}\sum_{n=0}^\infty e^{inx}$.}.
In the case of the trivial phase
$$
\lim_{s\to\infty}C_s(x,y)\to2\pi\delta_{2\pi}(x-y).
$$
Hence, the canonical phase has the sharpest kernel $\lim_{s\to\infty} C_s$
and the trivial phase kernel loses all phase information of states.
Note also that $C_s(0,0)\le(s+1)^2$ and only in the case of the canonical phase
$C_s(0,0)=(s+1)^2$ for all $s$.

\begin{theorem}\label{t}
Let $(\psi,\fii)\mapsto f_{\psi,\fii}$ be a phase observable. 
For any (Borel) $X\subseteq[0,2\pi)$ for which
$\int_X\d\theta<2\pi$,
$$
\frac{1}{2\pi}\int_X f_{\psi,\psi}(\theta)\d\theta<1
$$
for all unit vectors $\psi\in H^2$.
\end{theorem}

\begin{proof}
Let $\psi\in H^2$ be a unit vector.
It follows from Theorem 3 of \cite{Pe2} that
$(2\pi)^{-1}\int_X f_{\psi,\psi}(\theta)\d\theta=\<U^*\psi|E(X)U^*\psi\>
=(2\pi)^{-1}\int_X(\theta|\Phi\left(\kb{U^*\psi}{U^*\psi}\right)|\theta)\d\theta$
where $\Phi$ is a covariant trace-preserving operation.
Since $\Phi\left(\kb{U^*\psi}{U^*\psi}\right)
=\sum_{k=0}^\infty\lambda_k|\fii_k\>\<\fii_k|$,
$\lambda_k\in[0,1]$, $\fii_k\in\cal H_N$, $\|\fii_k\|=1$, $k\in\N$, 
$\sum_{k=0}^\infty\lambda_k=1$, we get
$(2\pi)^{-1}\int_X f_{\psi,\psi}(\theta)\d\theta=
\sum_{k=0}^\infty\lambda_k(2\pi)^{-1}\int_X|(\theta|\fii_k\>|^2\d\theta$.
From Proposition 8 of \cite{BuLaPeYl} it follows that
$(2\pi)^{-1}\int_X|(\theta|\fii_k\>|^2\d\theta<1$ for any (Borel) $X\subseteq[0,2\pi)$ for which
$\int_X\d\theta<2\pi$ and this completes the proof.
\end{proof}
Theorem \ref{t} shows that we cannot find a phase wave function
which is localized in some phase window $X$ (essentially) other than $[0,2\pi)$, 
that is, a phase observable cannot be localized.
Also this shows that any phase observable $E$ and the number $N$ are
probabilistically complementary \cite{BuLaPeYl}.

\begin{remark}\rm
The first moment operator 
$$
A:=\frac{1}{2\pi}\int_0^{2\pi}\theta|\theta)(\theta|\d\theta=
\pi I+\sum_{n\ne m=0}^\infty\frac{i}{m-n}\kb n m
$$ 
of the canonical  phase observable 
$$
X\mapsto\frac{1}{2\pi}\int_X|\theta)(\theta|\d\theta=
\sum_{n,m=0}^\infty\frac{1}{2\pi}\int_Xe^{i(n-m)\theta}\d\theta\kb n m
$$ 
acts on the Hardy class $H^2$ as
$$
\psi\mapsto PQ\psi
$$
where $(Q\psi)(\theta)=\theta\psi(\theta)$ and $P$ is projection from $L^2([0,2\pi))$ 
to $H^2$. Thus, $A$ is the Toeplitz phase operator suggested by
Garrison and Wong \cite{GaWo} and also Galindo \cite{Ga} and Mlak \cite{Ml}.
The canonical phase is (essentially) the only phase observable which has a covariant 
projection valued dilatation to $L^2([0,2\pi))$ \cite{CaViLaPe}. 
The dilatation is the spectral measure $X\mapsto\chi_X$ of the multiplication
operator $Q$.
\end{remark}

\subsection*{Acknowledgments.}
The author thanks Dr.\ Pekka J.\ Lahti for many discussions, and 
for the careful reading of the article.

\end{document}